\documentstyle[sprocl,psfig]{article}

\def\Journal#1#2#3#4{{#1} {\bf #2}, #3 (#4)}

\def\PRD{{\em Phys. Rev.} D}

\def\be{\begin{equation}}
\def\ee{\end{equation}}
\def\bea{\begin{eqnarray}}
\def\eea{\end{eqnarray}}
\def\go{\mathrel{\raise.3ex\hbox{$>$}\mkern-14mu\lower0.6ex\hbox{$\sim$}}}
\def\lo{\mathrel{\raise.3ex\hbox{$<$}\mkern-14mu\lower0.6ex\hbox{$\sim$}}}
\begin{document}
\title{CURRENT HIGH ENERGY EMISSION FROM BLACK HOLES}
\author{R. D. BLANDFORD}
\address{130-33 Caltech\\Pasadena, CA 91125, USA\\E-mail: rdb@caltech.edu} 
\maketitle\abstracts{Two related topics are discussed. 1. Accretion onto 
black holes at low and high (though not very high)
rates is believed to proceed adiabatically
({\em ie} non-radiatively). It is argued that the liberated energy
is carried off by an outflow, probably involving almost all of the gas that
is supplied. Two dimensional, fluid, accretion disks, in which mass, angular 
momentum and energy are transported to the disk surface, are summarized. 
It is conjectured that relatively minor changes are needed to 
describe magnetised disks. By contrast, the disk surface physics is argued 
to dictate the character of the outflow. 2. Ultrarelativistic jets
appear to be produced by active galactic nuclei (AGN), pulsars and $\gamma$-ray
bursts (GRB). In all three cases, it is argued that the power is 
generated electromagnetically by a magnetic rotator, (in a DC not 
AC form), and transported 
in this manner to the emission site. A model of a relativistically 
expanding electromagnetic shell is described and 
used to provide a simple model of a GRB
in which the $\gamma$-rays are produced by unstable electrical currents
flowing along the rotation axis. The shell drives a 
relativistic blast wave into the surrounding medium with a speed 
that varies with latitude and whose afterglow emission may 
exhibit achromatic breaks. Similar processes may be at work in non-relativistic
plerions like the Crab Nebula and, possibly, AGN jets. The observational 
implications of these two classes of model and the prospects for performing
instructive, numerical experiments to elucidate them further are briefly 
outlined.}
\section{Introduction}\label{sec:intro}
I would like to take literally the title of this meeting and
summarize two, related, theories of how black holes 
(and, to some extent, magnetised neutron stars) behave. The first, which
draws upon research carried out
with Mitch Begelman, is a particular viewpoint 
on how accretion proceeds at rates 
that range from low to high. The second,
which is summarised in greater depth 
elsewhere\cite{bla02}, involves collaboration with Max Lyutikov. It 
contains some new approaches to describing how electromagnetic Poynting 
flux is released by a black hole (or indeed any compact spinning, magnetised
body) and how it
propagates to a remote site from which observable radiation
is emitted. I should caution
the reader that these are both active and controversial 
research areas and there are many other points of view that
I shall not review. However, I will try to highlight one
or two areas where I think that the differences between models 
are sharply delineated and where I anticipate that progress is imminent. 
\section{Ondol Dynamics}
\label{sec:ondol} 
\subsection{Modes of Accretion}\label{subsec:accmodes}
Accretion
\footnote{Ondol is a special Korean system of underfloor heating.}
onto a black hole is approximately scale-free, in the sense
that the mode of the accretion is most strongly determined by the ratio of 
the mass {\em supply} rate to the Eddington rate ($\dot M_{{\rm E}}=
4\pi GM/\kappa_Tc$) and the product of the
angular velocity of the hole, $\Omega$, with its mass $M$ [Ferrarese;
henceforth, other contributors to these proceedings will be designated
using square brackets].
(It is assumed that the gas has sufficient angular momentum
to form a disk.) There will be some sensitivity to the mass  
through the effective temperature.
The mass supply rate is usually estimated by the Bondi rate
$\dot M_B\sim\pi r_B^2\rho_Bs_B\sim\dot M_E\tau_Bc/s_B$, where $\rho_B,s_B,
\tau_B$ are the density, sound speed and Thomson optical depth at the Bondi 
radius, $r_B\sim2GM/s_B^2$.
This should be true for black holes with masses that 
range from stellar values $\sim5$~M$_\odot$ to 
the many billion solar masses that characterize the holes that 
are thought to power quasars \cite{kro99} \cite{fra92}.

Four accretion states can be distinguished, although it is hard to be
quantitative without a better understanding of the underlying 
fluid mechanics and plasma physics.
\begin{itemize}
\item{\em Drought ($\dot M<<\dot M_E$)}
The accreting gas is {\em adiabatic}. That is to say it is unable to cool 
and radiate away its binding energy within some {\em transition} 
radius $r_{{\rm trans}}$. This requires that the 
internal energy be carried mostly by hot ions which do not heat the 
electrons enough to enable them to radiate efficiently on the inflow timescale
\cite{nar98}.
Instead, the surplus energy, which is transported outward by magnetic stress, 
is carried off in an outflow. This may account for most of the mass 
supply so that the hole accretion rate is very much less than the mass supply
\cite{bla99}.

Possible examples include low power, galactic nuclei,
(including Sgr A$^\ast$), AGN with powerful radio sources (including
quasars) and binary X-ray sources in their low hard states [Menou, Mineshige].
\item{\em Rain ($\dot M\sim\dot M_E$)}
The gas can radiate efficiently all the way down to the marginally stable
circular orbit $r_{{\rm ms}}$. The flow is conservative so that the accretion
rate is similar to $\dot M$
\cite{sha73}. Most of the radiation will be emitted 
quasi-thermally by the disk, although there is likely to be considerable, 
nonthermal, coronal activity as well. The inner disk is radiation- and electron
scattering-dominated with thickness $H\sim\dot M\kappa_T/4\pi c$. 
These disks are
expected and observed to be unstable and are probably quite inhomogeneous.
\footnote{This may permit radiation to escape with a luminosity in excess of 
the Eddington limit\cite{beg02}.}
The high gas density and the 
intense radiation field may prevent the formation of relativistic
jets.

Possible examples include Seyfert galaxies, radio-quiet quasars 
and binary X-ray sources in their high, soft states.
\item{\em Deluge ($\dot M_E\lo\dot M\lo\dot M_E(c/s_B)^2$)}
Radiation is trapped within the trapping radius 
$r_{{\rm trap}}\sim\dot M\kappa_T/4\pi c$ within which the accreting gas 
flows faster than the photons can diffuse. 
The flow is again adiabatic and the energy that is released is carried
off by a mildly relativistic, radiation-dominated wind 
\cite{bla99} [Park]. The hole accretes at the roughly the Eddington rate. 

Possible examples include broad absorption line quasars and Galactic 
superluminal sources like GRS1915+105. 
\item{\em Inundation ($\dot M\go \dot M_E(c/s_B)^2$)}
The trapping radius lies beyond $r_B$. Strongly super-Eddington accretion
is possible because outflows from the inner disk are unable to escape.

Possible examples include massive black holes
during their rapid growth phases and some GRB models.
\end{itemize} 

The reason why outflows are inevitable when the flow is 
adiabatic is that
the torque $G$ in the disk automatically transports energy
radially outward at a rate $G\Omega$, where $\Omega$ is the angular
velcocity \cite{bla99}. There is a local heating rate
$G\nabla\Omega$ which can unbind gas in a nearly Keplerian disk.
\footnote{This is not necessarily true for a disk that extends beyond 
$r_B$ where the potential is much softer.} 
The energy liberated by each proton that (altruistically) cross the horizon
drives a powerful wind from larger radii which allows many
(up to $10^5$) other protons to escape. 
\footnote{It is possible that the energy is carried off hydromagnetically, 
in which case, there need be little mass loss.} 
\subsection{The Importance of Magnetic Field}\label{subsec:impmag}
Early models of accretion disks \cite{sha73} were non-specific
about the nature of the viscous torque allowing it to be either 
fluid or hydromagnetic in character. We now know that, under almost 
all circumstances, the latter is correct. The magnetorotational
instability \cite{bal98} ensures that magnetic fields grow to dynamically 
important strength on the orbital timescale [Vishniac]. 
The nonlinear evolution
of this instability is still far from clear. However, it is likely
that it involves coronal heating, jet formation, and possibly
emission line cloud dynamics.

What is clear is that we are still quite ignorant of the true laws
of MHD. The global evolution of magnetic field, as witnessed 
much closer to home in the terrestrial magnetosphere and the 
solar corona, is controlled by principles of global stability 
and physical processes like reconnection, conduction and equilibration, 
which we do not understand well enough 
to generalize to astrophysical environments. 
However, the prospects are generally quite good 
for improving our understanding of these matters through more detailed
space physical observation, terrestrial experiments using giant 
pinches, lasers and particle beams as well as numerical simulation.  
\subsection{Two Dimensional Adiabatic Disks}\label{subsec:accdisk}
Despite these remarks it is still instructive (and easier) 
to consider fluid disks. These are inevitably at least two dimensional
[Lee]. 
A useful approximation is that the viscosity is 
small enough that they can be treated as being in approximate hydrostatic 
equilibrium. This requires specifying how the density varies along isobars
which, in turn, depends upon the manner in which energy, mass and angular 
momentum are transported through the disk.

Two dimensional adiabatic disks are likely to become linearly unstable 
according to the H\o iland criterion, which is a linear combination of the 
more familiar Schwarzschild and Rayleigh prescriptions for instability
and appropriate for hot, rotating flows. There are actually two criteria. Most 
accretion disks are quite stable to the first of these which essentially refers 
to radial interchanges. It is the second criterion that is more relevant.
It essentially states that fluid disks are marginally unstable 
to the interchange of slender rings when they are 
gyrentropic; that is to say when the surfaces of constant entropy and specific 
angular momentum coincide. (In fact this implies that the surfaces 
of constant Bernoulli function are also coincident.) Furthermore, the 
nature of the unstable modes is such that if a disk is only slightly unstable,
the motions of fluid elements will transport mass, angular momentum and 
energy along these gyrentropes to the disk surface where they 
can be transferred to an outflow. 

The precise properties of the outflow depend upon the detailed nature
of the dissipation and momentum transfer that takes place at the 
disk surface and these processes are not even understood in the case of 
the sun. If the wind is thermally driven, then there 
must be an entropy jump at its base as the gas in the disk changes from 
being bound to unbound. However it also possible that the gas remains
relatively cold and that magnetic stress create a momentum-driven 
wind as appears to happens in the case of the solar wind, 
where two million degree gas can
acquire a velocity at infinity of $\sim800$~km s$^{-1}$.
In addition, there is the possibility of launching 
a cold, magnetocentrifugal wind from a near Keplerian accretion disk
-- a possibility that does not exist in the case of the sun. The 
key conclusion is that the structure
of adiabatic, accretion disks depends upon the unknown physics of mass 
outflow. 

However, even this is not the whole story. The gas that remains
in the disk must flow 
inward on a viscous timescale and the manner in which this happens
depends upon the prescription for the viscosity. Indeed, in the presence of 
convective motion, there is a natural quadrupolar 
circulation in the disk which will be 
established and this must be superposed upon the net flow to the disk 
surface and smaller radii. Just as 
is the case with Eddington-Sweet circulation,
the flow adjusts to ensure that there is no local accumulation
of mass angular momentum and energy. 
\footnote{If we do not admit the possibility of an outflow from the disk 
surface, then the global circulation will transport heat from 
small radii, where it is mostly liberated, to $r_{{\rm trans}}$ 
where it must either
be radiated away or continually inflate the disk so that the flow becomes 
unsteady.} 

It is possible to construct self-consistent, self-similar models that exhibit
convection, inflow, outflow and circulation after specifying a functional
form for the viscous torque and outflow launching mechanism\cite{bla02}. 
These models
are generic and their general features are displayed by 
some numerical simulations 
of two dimensional disks\cite{sto99}.
Self-similar disks are subject to the criticism that they cannot describe
the flow close to the hole, where most of the energy is released. This
criticism can be met by constructing an explicit non self-similar, 
relativistic torus model around $r_{{\rm ms}}$. The physical assumptions
that must be introduced to create such a model are necessarily arbitrary 
but they suffice to demonstrate that there is no difficulty of principle 
associated with terminating a self-similar solution at small radius.
Similar concerns have been raised concerning the outer radius. 
Understanding this transition
has turned out to be a much harder problem.   
\subsection{Magnetized Disks}
We have just argued that fluid disks are gyrentropic. However, this is 
unlikely to be true of magnetised disks. The dynamics  
is clearly complex [Krolik]. For example, 
if it turns out that the magnetic flux 
preferentially settles into isorotational surfaces and distributes 
the angular momentum rapidly on these surfaces, then the disk 
structure will be barytropic. This prescription has some numerical support
\cite{sto01}.
Whatever the correct prescription, the self-similar models 
described above can be modifed with only minor changes.
It is the prescription for generating the outflow
at the disk surface that is more important for determining the disk 
structure.
\section{Seungmu Dynamics}
\label{sec:seungmu}
\subsection{Ultrarelativistic Jets}\label{subsec:ultrarel}
Ultrarelativistic 
\footnote{Seungmu dancers are able to beat a drum and, simultaneously,
spin long tassles attached to their hats.}
jets are found in a variety of locales.
They were first seen in extragalactic radio sources
[Celotti, Sikora]. Indirect arguments,
together with direct measurement of superluminal expansion show
that the emitting material can move with a Lorentz factor
$\Gamma\sim10$ and be collimated into a cone with opening
angle $\sim5^\circ$. There is now evidence that  
their sources are spinning, massive black holes with 
their attendant accretion disks and that they are collimated
within $\sim100m$. Most jets appear to be associated 
with accretion disks that radiate well below their Eddington limits and
carry as much, or even more, power than is radiated by the 
disk.
\footnote{It is well worth searching for the stellar counterparts
of ultrarelativistic jets  - Galactic blazars that are 
beamed towards us.}

The second example of ultrarelativistic jets is provided by plerions. These
are supernova remnants, exempified by the Crab Nebula, with central, 
rapidly spinning, magnetised, neutron stars. The rotational energy
of the neutron star appears to be carried off by a relativistic 
outflow. Recent, observations by the Chandra Observatory
(now replicated in other plerions)
show a pair of jet-like features, together with an equatorial disk of X-ray
emission. This was surprising because it was thought that jets
required an accretion disk to form. The region, that is observed directly,
is clearly not moving with ultrarelativistic speed 
as they would then be beamed away from 
us. However, it is a reasonable supposition that they contain invisible,
ultrarelativistic cores.   

The third example is provided by GRBs which are known to be
cosmologically distant and, consequently, extremely energetic.
It has been inferred that the outflow is ultrarelativistic,
with Lorentz factors $\Gamma\sim300$ as $\gamma$-rays
with energy in excess of $\sim0.5$~MeV have to escape without creating pairs
\cite{mes02}.
They are popularly associated with stellar compact objects and it 
was guessed that they might be beamed, in order
that the burst energies not be unreasonably large.  
Reports of achromatic breaks in afterglow spectra
from some long duration bursts,
which can be formed when $\Gamma$ decreases to of order the
reciprocal of the jet opening angle, support this view.
\subsection{Magnetic Rotators}\label{subsec:rotat}
A reasonable guess is that all of these ultrarelativistic outflows
(if not the more general class of cosmic jets) are due to similar physical 
processes. However, although popular models of AGN and plerionic jets are
essentially electro- or hydromagnetic, most GRB models invoke 
a ``hot big bang'' that produces a fireball with a very high entropy 
per baryon, just like the early universe. I would like to explore the 
alternative hypothesis that all three types of source are really quite similar 
and derive their power from the continuous
extraction of rotational energy from a compact
object by electromagnetic stress - a ``cold, steady
state'' model instead of a hot, big bang!
\footnote{It is amusing that, if we are correct in identifying
long duration GRBs with compact objects of size $\sim10$~km,
then their sources are observed to be active for a million light crossing
times -- an order of magnitude greater than the number of crossing
times that we have observed a typical quasar!} 
Furthermore, I suppose that 
the energy remains mostly in an electromagnetic form
as it is transported into the emission region.

This magnetic rotator model 
is best developed in the case of plerions, where it is supposed
that the central spinning neutron star possesses an inclined dipole 
moment and that it is surrounded by a force-free magnetosphere through which
currents flow and space charge is maintained [Hirotani]. 
The complete electrodynamical
description of this magnetosphere remains an unsolved problem. However,
it seems likely that somewhere beyond the light cylinder, the electromagnetic
field becomes essentially axisymmetric and that variation on the scale of a
wavelength dies away. There are at least three ways by which this can occur.
There can be steady reconnection in the outflowing, ``striped'' wind.
Alternatively, the waves can decay through parametric instability
into higher frequency waves. These two processes are essentially 
dissipative. Finally, the minority of magnetic field lines
that emanate from the neutron star's southern
magnetic pole, and which can be traced into the northern hemisphere,
may gradually be pulled by magnetic tension across the equatorial 
plane back into the southern hemisphere
(and {\em vice versa}). This can happen non-dissipatively
near the light cylinder. In summary,
I will presume that only the DC, not the AC component of the electromagnetic
field survives. 

If this simplification of the magnetic geometry actually takes place, 
we will be left with a relativistic wind containing a predominantly 
toroidal magnetic field spun off by the central body.
Associated with this toroidal magnetic field will be a poloidal
electrical field that is almost equal in magnitude.
(Many authors have argued that this wind quickly becomes 
plasma-dominated and terminates in a relativistic, fluid shock front 
in the inner nebula. However, it is very hard to see how 
the DC magnetic field can be erased so quickly and, 
although there are moving ``wisp'' features, these
appear to lie in the equatorial plane.
There is really no evidence for a relativistic shock in tne nebula.)

There are several suggestions as to how the jets associated with 
extragalactic radio sources are launched. One of the simplest
is that the power for the ultrarelativistic outflow is extracted 
as Poynting flux from the spin of the hole by magnetic field lines 
that are supported by external currents flowing in the disk. 
It is also possible to extract energy electromagnetically 
from a surrounding accretion disk assuming that enough 
of its area is threaded by open magnetic field. However, this is most 
likely to produce a hydromagnetic outflow where 
the terminal velocity is no more 
than mildly relativistic although it may be  responsible
for collimating the much faster flow from the black hole.

When we consider electromagnetic models of GRBs,
we find several advantages. The most fundamental
is that the electromagnetic stress tensor, is anisotropic, in contrast to 
an isotropic fluid pressure tensor. This, implies that electromagnetic
outflows can be naturally self-collimating.
In addition, the presence of a dominant electromagnetic field implies that
the effective internal sound speed is that of the fast mode which 
is close to the speed of light. Electromagnetic jets, in contrast
to fluid jets are no more than mildly hypersonic.
\footnote{Contrast this with the hypernova model of GRBs
\cite{mes02} [Maeda, Lee]
where it is supposed that a high entropy per baryon fluid is collimated 
by a vortex inside a star so that the pairs and $\gamma$-rays eventually
transfer their energy to protons and the ratio
of the momentum flux to the pressure is $>3\Gamma^2\sim3\times 10^5$. 
This seems very unlikely to be true of a naturally occuring explosion, 
especially in the region where internal shocks are supposed to be operating.}

They are few direct clues as to the prime movers of GRBs
\cite{mes02}. 
The closest model to that of AGN jets   
has a spinning, stellar black hole surrounding by a stellar 
mass torus - possibly a tidally destroyed neutron star - that 
confines a $\sim10^{14}$~G magnetic field
[van Putten]. The emission lasts as long as 
the torus survives. The model that is closest to plerions involves
a rapidly-spinning magnetar that has just been formed by 
accretion-induced collapse of a white dwarf. Alternatively, it might
be possible for an accreting magnetised neutron star that is 
collapsing under rotational support to form a black hole or for two 
merging pulsars to act as magnetic rotators
\footnote{In principle, it is possible that an electromagnetic jet can 
be formed inside a collapsing star, though the star is not needed to 
provide the collimation and it seems
hard to believe that plasma can be excluded from the outflow as efficiently
as required.}  

The common feature of all of these putative sources is that they spin 
off a toroidal magnetic field and an associated, electromagnetic 
Poynting flux that is unburdened by baryons.
\subsection{Electromagnetic Black Holes}\label{subsec:electro}
The idea that the spin energy of a black hole can be extracted
electromagnetically has received an observational boost from the discovery 
that black holes are commonplace on both the stellar and the massive scale
(as well as, perhaps, on the intermediate scale) and that the second parameter,
the spin is large so as to allow gas to orbit close to the horizon and to
form strongly redshifted emission lines
\cite{wil01}. 

There are several ways through
which the rotational energy associated with the spinning spacetime can be 
tapped electromagnetically
\cite{bla01}. The particular choice that I have emphasised,
because I believe that it represents the dominant energy channel, is that
the horizon is threaded by a large flux of open magnetic field [Park]. 
A continuous,
electromagnetic Penrose process operates in the ergosphere of the black hole
which results in Poynting flux flowing inward across the horizon 
and, simultaneously, propagating away from the hole to infinity. 
\footnote{The energy flow is 
conserved in Boyer-Lindquist coordinates 
and so power appears to emerge from the 
horizon in the Boyer-Lindquist frame. 
However, physical observers must orbit with respect 
to this coordinate system. Doppler boosting the energy flux into a frame 
moving with a physical observer produces an inwardly directed energy flux.} 
The source of the power is ultimately the reducible mass
of the hole, from which the electromagnetic field in the ergosphere
is extracting work.

However, not all the field lines that thread the horizon need connect with the
outflow. Some low latitude field lines may connect directly to the accretion 
disk and provide a supplementary power source for the disk as well as 
a possible driving torque for exciting quasi-periodic oscillations [Dotani]. 
This energy channel could be important,
especially if the disk is thick. However, it is unlikely to lead to 
an ultrarelativistic outflow. 
\footnote{The magnetic connection of the disk to the 
plunging gas seems to be a less promising source of 
power because the magnetic field lines quickly reconnect leaving 
the gas effectively disconnected from the disk \cite{ago01}.}

The process that I have just described is distinct from (though can operate
simultaneously with) an alternative process, the extraction of binding energy 
by open field lines
from the accreting gas both in the disk and in the plunging region
between the inner edge of the disk and the horizon
\cite{mei01}. The extra power 
that this process produces can be 
charged to the spin energy of the hole,
which increases at a slower rate than it would do so in the absence 
of magnetic stress. However
the intermediate working substance that effects this transformation
is the inertia of the plasma not the electromagnetic field. 

There are three 
reasons for emphasising direct extraction of energy from the
hole to extraction from the infalling gas at least for a rapidly 
spinning hole. The first is that the event horizon has a larger effective
area than the annular ring between the hole and the disk. The second is that
any gas-driven outflow is likely to be contaminated with 
baryons and consequently
is unlikely to achieve an ultrarelativistic outflow velocity
required.  The third is that holes probably rotate
much faster than orbiting gas, except quite close to the horizon, from where
the extraction of energy will be quite inefficient.
\subsection{Electromagnetic Shells}\label{subsec:shells}
Suppose that a magnetic rotator spins off
magnetic flux into the far field for a time $t_{{\rm source}}$
and that this creates a relativistically
expanding shell of electromagnetic field, of thickness $ct_{{\rm source}}$ 
that drives a blast wave into the surrounding medium
(Fig.~1). The blast wave is
supposed to be bounded on its outside by a strong shock front that moves with 
Lorentz factor $\Gamma$ and, on its inside, by a contact discontinuity,
separating it from the shell, that moves with Lorentz factor $\Gamma_c$.   
Suppose further, for simplicity, that the current  
well beyond the light cylinder flows along the axes, then
along the contact discontinuity and finally  returns 
to it source through the equatorial plane. 
\begin{figure}[htb!]
\hfil
\psfig{figure=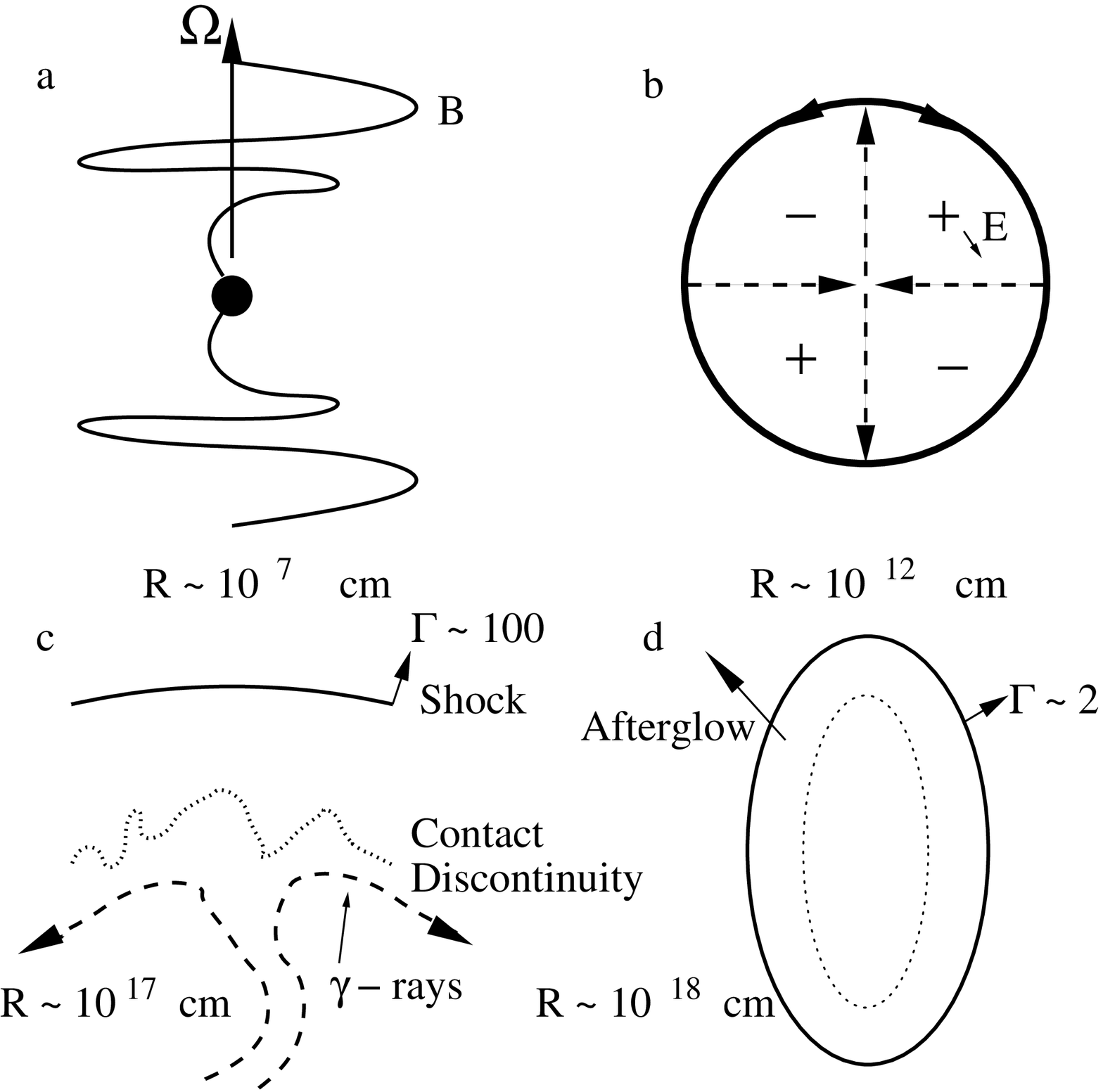,width=0.8\textwidth}
\hfil
\caption{Four stages in the expansion of a magnetic shell with scales
appropriate to a long duration GRB.
a) The magnetic field $\vec B$ changes from poloidal to
toroidal close to the outgoing light surface of the magnetic rotator
at a radius $R\sim10^6$~cm. The alternating component of the electromagnetic
field decays relative to the DC toroidal field. b) The source is active 
for $\sim100$~s. By this time, it will have inflated a magnetic
bubble with radius $R\sim3\times10^{12}$~cm, expanding with Lorentz factor
$\Gamma\sim3\times10^4$.The magnetic field is mostly toroidal, with the
signs shown, while the electric field $\vec E$ is poloidal. The quadrupolar 
current flow is shown dashed. The shocked circumstellar medium
is compresed into a thin shell of thickness $\sim10^3$~cm.
c) By the time the shell has expanded to $R\sim10^{17}$~cm 
$\Gamma\sim100$ and most of the 
electromagnetic pulse has caught up with the blast wave. This phase
is observed a time $\sim100$~s after the initial explosion. The current flow
is still largely quadrupolar, though it is unstable along the axis 
and the equator and this drives an electromagnetic turbulent cascade.
which ultimately creates electrical resistance and dissipation in the form
of pair production, particle acceleration and intermittent, 
$\gamma$-ray emission. These instablities also promote corrugation of the
contact discontinuity and incorporation of the magnetic field into
the shocked interstellar medium where it can mix with relativistic electrons
accelerated at the bounding shock front. d) When the blast wave has expanded
by a further factor ten, its speed is only mildly relativistic. Its shape
will be quite prolate as the expansion is fastest along the axis. Most of 
the energy released by the central, spinning, magnetic rotator is now carried
by the shocked interstellar medium.
\label{fig:magshell}}
\end{figure}
If we ignore the poloidal component of the magnetic field, (and
consequently the flux of angular momentum),
then the relevant solution of the force-free equations [Park] associated 
with this current flow has the form 
\bea
\label{shell}
B_\phi&=&{f_+(t-r)+f_-(t+r)\over r\sin\theta}\\
E_\theta&=&{f_+(t-r)-f_-(t+r)\over r\sin\theta}
\eea  
The two terms in each expression are fast modes propagating outward and inward.
The charge and current density vanish in the interior
of the shell. We can determine the functions 
$f_+,f_-$ by specifying the electromagnetic field
at some small radius beyond the light surface and by matching to an 
ultrarelativistic blast wave expanding into the surrounding medium. 
This last requires that the outer surface of the shell move at the same 
speed as the inner surface of the blast wave and that the magnetic stress 
normal to this surface match the pressure in the blast wave.

The simplest assumption to make is that the
strength of the magnetic rotator is constant, ($f_+={\rm const}$) for a 
time $t_{{\rm source}}$ and that the external
density is constant in radius. These assumptions imply that the 
Lorentz factor of the blast wave's outer shock front varies with radius 
$R$ according to $\Gamma\propto\csc\theta R^{-1/2}$
\cite{bla76}. 
(The
Lorentz factor of the contact discontinuity, $\Gamma_c$, exhibits a similar
variation.) The 
electromagnetic velocity (the velocity of the frame in which the 
electric field vanishes) in the body of the shell
$\vec\beta=\vec E\times\vec B/B^2$ is radial and equal in magnitude
to $(f_+-f_-)/(f_++f_-)$ and the magnetic stress in a frame
moving with this velocity is $\propto f_+f_-\csc^2\theta$.
The expansion of the blast wave 
is anisotropic, being faster along the poles,
giving an electromagnetic power per steradian $L_\Omega\propto\csc^2\theta$.
Note that information is propagated inward by the fast mode which does
modify the solution and allows it to react to changes in the load including 
changes in its effective impedance. Note also that it takes a
very long time for a wave to be reflected by the blast wave and return
to the origin. This is generally true of ultrarelativistic flows
and stationary solutions, which take a long time to be established
can be quite misleading.

A comparison with a non-relativistic plerion
like the Crab Nebula, is instructive. The magnetic 
bubble expands with speed of $\dot R<<c$
and, consequently, the pulsar must be producing 
magnetic flux at a rate that is roughly $(c/\dot R)^{1/2}$ 
times too large to account for the strength of the magnetic 
field in the nebula. Therefore, most (95 percent in the case of the 
Crab Nebula) of the flux 
must be destroyed. On topological grounds, the natural
places for this destruction to occur are on the axis and the equatorial
plane. These regions are, in any case, 
formally unstable to pinch and tearing mode instabilities
\footnote{The contact discontinuity should be formally
Kruskal-Schwarzschild stable.}
respectively
\cite{beg98}. 
\footnote{If we were to prevent this dissipation, in a thought experiment,
through having the current flow along rigid, perfect conductors, then
the reflected electromagnetic waves would react back on the source and
shut down the power supply.}
The X-ray emission from the Crab Nebula, observed 
by Chandra
\cite{wei00}
, may well be a consequence of particle
acceleration associated with electrical resistance / 
flux destruction / generation of electromagnetic turbulence / 
particle acceleration tracing out the current flow 
in the inner part of the nebula. If we consider the flow of 
electromagnetic field in the nebula, we find that there 
must be a steady flow of magnetic flux towards the rotation axis and the 
equatorial plane
\cite{beg98}. In addition, there will be a compensatory backflow of 
relativistic electrons and positrons into the interior of the plerion.
Of course there must be some matter in the nebula -- it has been 
observed evaporating off filaments through its polarisation behaviour --
but in this model, it has a very minor role in the dynamics.

By contrast, the contact discontinuity of an electromagnetic shell
expands at a relativistic speed, and the reflected wave has a 
much smaller amplitude
than the incident wave. Consequently, there is no necessity 
to destroy a lot of magnetic flux.
Stated another way, there {\em need} be little resistance in 
the circuit. The effective load 
consists of the performance of work on the expanding blast wave. This is 
where most of the power that is generated by the central magnetic
rotator ends up. (The distinction between this inertial load and a 
dissipative load is quite unimportant for the behavior of the 
black hole magnetosphere.) 

The simple electromagnetic solution will only remain valid until the end of
the  
outward-propagating, electromagnetic pulse catches up with the blast wave.
This occurs at some radius $R_{{\rm free}}\sim\Gamma(R_{{\rm free}})^2
ct_{{\rm source}}$. Thereafter, the surrounding blast wave which, by now,
has acquired almost all of the energy in the explosion, will expand freely 
with $\Gamma\propto R^{-3/2}$.

This elementary model of a relativistic electromagnetic shell
can be easily generalized to accommodate different assumptions
about the variation of the flux supply with time and latitude
and the density variation in the external medium.  
\subsection{Gamma Ray Bursts}\label{subsec:grbs} 
As well as bring out some formal points, the electromagnetic solution
just described provides a possible model for GRB
\cite{uso94} \cite{lyu01}.
Suppose that a magnetic rotator spins off toroidal magnetic 
field as it slows down and that this magnetic field fills an 
anisotropic, relativistically expanding shell in a uniform medium.
Suppose, further, that the flux 
distribution near the light cylinder is consistent 
with the current being concentrated along the axis and in the equatorial
plane, as described above. The current density is most intense 
on the axis and, although there is no requirement that flux be destroyed as 
long as the expansion is relativistic, in practice the magnetic pinch will
become hydromagnetically unstable 
to sausage and kink modes (in the comoving frame) after expansion 
beyond a radius where the stabilising, poloidal field becomes insignificant
\cite{beg98}.
These global instabilities, which should have a longitudinal wavelength 
comparable in size to the width of the current distribution, 
may sustain an electromagnetic turbulence spectrum which should 
ultimately be responsible for particle  
acceleration and the excitation of transverse gyrational motion
\cite{tho98}. 
\footnote{This turbulence 
may have already been seen in the measured fluctuation power spectrum
[Chang].}
The reason why particles are accelerated is that, when the power cascades
down to short enough wavelengths, there are too few charged particles 
to carry the electrical current and field-parallel electric fields will 
develop. This is the microscopic source of the dissipation.
These particles will emit $\gamma$-rays, primarily through 
the synchrotron process, though inverse Compton scattering
may also be important.
$\gamma$-rays, in excess of threshold
will create fresh pairs. All of this will take place in a frame moving
with the electromagnetic velocity and the emission will be 
strongly beamed outward. As well as create
electrical resistance, the global pinch instabilities can also provide a 
plausible explanation for the large $\sim1-10$~ms fluctuations in the observed 
$\gamma$-ray flux that are observed.
\footnote{Attributing this variation to 
internal shocks tied to the source, as in the fluid model,
implies that the $\gamma$-rays originate at much smaller radii than 
expected on the electromagnetic model.}
 
The afterglow is formed after the blast wave becomes free of
its electromagnetic driver. Now, in most afterglow models, 
including those involving jets, it is supposed that the 
expansion velocity does not vary with latitude. However,
an electromagnetically-driven 
blast wave necessarily creates an anisotropic explosion and this has important
consequences for observations of the afterglow, especially in the 
ultrarelativistic phase of expansion. If we continue to use our simple model,
we find that the afterglow expansion varies most rapidly, and remains
relativistic for longest, closest to the symmetry axis. 
As $L_\Omega\propto\csc^2\theta$ the energy contained in each 
octave of $\theta$ is roughly constant
This means that the most intense bursts and afterglows 
in a flux-limited sample will be seen pole-on 
and should  exhibit achromatic breaks, which might be mistaken for jets.
The inferred explosion energy will be roughly independent
of $\theta$ and characteristic of the total energy.
When the expansion becomes non-relativistic, the remnant will have a 
prolate shape which might be measurable.
\footnote{It is tempting to associate some of the barrel-shaped 
supernova remnants observed in our Galaxy with the remnants of 
electromagnetic explosions.} 

This electromagnetic model provides a solution to the puzzle
of how to launch a blast wave that extends over an angular scale
$>>\Gamma^{-1}$ and where the individual parts are out of causal contact.
In the electromagnetic model, the energy is transferred to the blast wave
by a magnetic shell that pushes (unevenly) on the surrounding gas
all the way out to $R_{{\rm free}}$. It also supplies an origin
for the magnetic flux in the blast wave, for which the alternative
origin in the bounding shock front seems very hard to explain. 
In the present model, magnetic field can simply 
be mixed into the blast wave (and the shock-accelerated
relativistic electrons) at the contact discontinuity
through instabilities, much like what seems to happen in regular supernova
remnants.
\subsection{Some Numbers}\label{subsec:numbers}
Let us give some illustrative orders of magnitude for a model of a long 
duration GRB. The electromagnetic energy flux near the pole is 
$L_\Omega\sim10^{50}$~erg s$^{-1}$ sterad$^{-1}$ and lasts for 
a time $t_{{\rm source}}\sim100$~s. The associated EMF in the electrical
circuit $\sim10^{22}$~V 
\footnote{A potential difference this large, made available along the 
contact discontinuity, provides one of the few astrophysical options for 
accounting for UHE cosmic rays.} 
and the current is $\sim10^{20}$~A. 
\footnote{For comparison the values are $\sim3\times10^{14}$~A, $\sim3\times
10^{16}$~V for the Crab Nebula and $\sim10^{18}$~A, $\sim10^{20}$~V
for Cygnus A.}
The external density is uniform 
$n\sim1$~cm$^{-3}$. The blast wave is driven by the electromagnetic shell 
with Lorentz factor $\Gamma\propto R^{-1/2}$ until $R\sim R_{{\rm free}}
\sim10^{17}$~cm, $\Gamma\sim\Gamma_{{\rm free}}\sim100$. 
Thereafter there is a freely expanding
blast wave with $\Gamma\propto R^{-3/2}$ until the expansion becomes 
non-relativistic when $R\sim R_{{\rm NR}}\sim3\times10^{18}$~cm.

Most of the GRB emission  (around $\sim1$~MeV)
is produced when $R\lo R_{{\rm free}}$ as synchrotron emission
by $\sim100$~GeV electrons in a co-moving magnetic field of strength 
$B\go30$~G. The comoving cooling time of these electrons is $\sim3$~s,
a fraction $\lo10^{-4}$ of the expansion timescale and so
if the $\sim100$~GeV pair energy density is maintained at a 
significant fraction of the equipartition energy density, then the magnetic
energy can be efficiently transformed into $\gamma$-rays. 
The opacity to pair production for a $\gamma$-ray of energy $E$
is $\sim0.1(E/1{{\rm GeV}})$. The Thomson optical depth depends upon 
the details of the particle acceleration but is 
plausibly much smaller than unity so that the observed $\gamma$-rays can
freely escape without erasing the variability.   
\subsection{AGN Jets}\label{subsec:jets}
Having briefly discussed electromagnetic models of plerions and 
GRBs, it remains to re-consider extragalactic jets in this context
[Fletcher]. Although
electromagnetic/hydromagnetic models of the energy release
have been quite widely discussed for a long while, it has generally been 
supposed that this electromagnetic energy is transformed into a 
particle-dominated flow at a safe distance from the black hole and that the 
observed emission is from a high $\beta$ plasma. 

How plausible is it that
the entire radio source is an electromagnetic structure,
that the observed emission trace out 
unstable currents rather than strong shock fronts?
\footnote{Of course, under force-free conditions, the currents also trace 
out the magnetic field in the frame in which the electric field vanishes.}
One of the merits of this
hypothesis is that it may provide a dynamical rationalisation of the 
doctrine of equipartition. Pinched currents, on all scales, may continue
to become unstable until their stresses are balanced by pressure. 
Another merit is that it provides a 
natural explanation for the helical structures that are often seen
in VLBI maps. A third advantage is that currents can account
for distributed  particle acceleration in well-resolved jets, as spectral 
studies suggest may be required. However, if
the extended lobes associated with 
the powerful FRII sources, like Cygnus A, are filled with unstable
though fundamentally toroidal and force- free magnetic 
field, then this could present a quite strong signal in the Faraday 
rotation maps which has not really been seen. Similarly the polarisation 
structure of FRII radio sources does look like a shear flow and there
are some features, in radio maps, like the knots in M87,
which are naturally interpreted as strong shocks.
\section{Summary and Prospects}\label{sec:summary}
In these impressionistic sketches, I have outlined two principles, which 
although not completely new, are now not commonly discussed -- 
that most accreting
black holes (excluding those where the mass supply is within an order
of magnitude or so of the
Eddington rate) swallow only a tiny fraction of the 
gas supplied at the Bondi radius and that ultrarelativistic,
high energy phenomena are fundamentally electromagnetic not 
gas dynamical. The link between these two principles may 
be that carrying off the binding energy of the accreting 
gas in a fluid outflow, rather than
radiation, is a necessary condition for extracting energy 
electromagnetically from a spinning black hole. At least this is the story
in the case of AGN jets. This condition is also satisfied by
pulsars. GRBs, which
may be either accreting black holes or neutron stars, would be similar 
to either. An extension of this line of argument is that AGN jets may 
be essentially electromagnetic all the way to their hot spots.

There are many possible discriminatory observations. For the most interesting 
and immediate case of GRBs, it is not predicted that there will be an 
accompanying neutrino signal. By contrast, 
a gravitational wave signal is expected 
in some, though not all, 
models and would be strongly diagnostic if detected. It is unlikely, 
though not impossible, that GRBs will be associated with Type II supernovae.
More immediate but less specific diagnostics include relating the 
duration and character of the GRB with the inferred 
observation angle of the burst -- higher inclination should be associated
with longer and less intense bursts. The spectrum and polarisation of the 
afterglow emission might also contain some clues, though the lack of 
a usable theory of particle acceleration 
\footnote{Recent, promising progress on understanding 
relativistic shocks predicts
a power law distribution function with a logarithmic slope of 2.2
\cite{kir00},
provided that the scattering is essentially normal to the shock front.
What is not yet clear is whether these scattering conditions are present.}
and magnetic field amplification at ultrarelativistic shocks
makes this a bit problematic. A more detailed discussion 
of the GRB emission, than presented
here should account for the MeV breaks observed in $\gamma$-ray spectra.

From a more theoretical 
perspective, there is much to be learned about the properties of 
force-free electromagnetic fields and especially their stablity. The 
possible relationship of the GRB fluctuation power spectrum to an 
underlying turbulence spectrum is especially tantalising. Undoubtedly,
numerical simulations will be crucial as the problem is essentially
three dimensional. Force-free electromagnetism is easier to study
than relativistic MHD and may well be a very good approximation
in many of these sources.

Turning to plerions, the most direct, observational challenge is to see
if there really is a strong, dissipating shock as expected with a fluid wind
or a flow of electromagnetic energy towards the axis and the equatorial
plane as predicted by the electromagnetic model and as appears to 
be exhibited by the Crab Nebula.

Finally, for AGN, we would like to detect and understand the mass outflows
predicted above for low mass supply rates. (We already know that rapidly 
accreting stellar and massive holes drive dense, high speed outflows.) 
In addition, we need to see if the very well-observed 
radio jets and their lobes can be re-interpreted in terms of an 
unstable Z-pinch. A good place to start is through mapping the 
magnetic field using polarisation measurements and the
internal mass density through internal depolarisation data.
In addition, detailed imaging spectra
from radio to X-ray energies can be used to determine where the 
particles are being accelerated -- at shock surfaces or in volumes
containing strong, unstable currents -- and how they propagate away from
these acceleration sites at different energies.

The conjunction of recent, impressive discoveries, 
upcoming observational facilities and powerful computing capability 
makes this a propitious time to be 
studying current high energy emission from black holes.
\section*{Acknowledgments}
I thank the organisers of this workshop for their gracious and generous
hospitality, my colleagues Mitch Begelman and Max Lyutikov for their 
collaboration on the above and several
of the conference participants for discussion which helped to clarify 
some of the arguments I have presented.
Support under NASA grant 5-2837 is gratefully acknowledged. 

\end{document}